\documentclass[12pt]{article}
\usepackage{amsfonts}
\usepackage{amssymb}
\usepackage{graphics,psboxit,amsmath}
\usepackage{subfigure}
\usepackage{graphicx}
\usepackage{verbatim}

\def\hybrid{\topmargin -30pt    \oddsidemargin 0pt 
        \headheight 0pt \headsep 0pt
        \textwidth 6.25in       
        \textheight 9.5in       
        \marginparwidth .875in
        \parskip 5pt plus 1pt   \jot = 1.5ex}

\hybrid

\def\baselinestretch{1.2}

\catcode`\@=11

\def\marginnote#1{}
\newcount\hour
\newcount\minute
\newtoks\amorpm
\hour=\time\divide\hour by60
\minute=\time{\multiply\hour by60 \global\advance\minute by-\hour}
\edef\standardtime{{\ifnum\hour<12 \global\amorpm={am}%
        \else\global\amorpm={pm}\advance\hour by-12 \fi
        \ifnum\hour=0 \hour=12 \fi
        \number\hour:\ifnum\minute<10 0\fi\number\minute\the\amorpm}}
\edef\militarytime{\number\hour:\ifnum\minute<10 0\fi\number\minute}

\def\draftlabel#1{{\@bsphack\if@filesw {\let\thepage\relax
   \xdef\@gtempa{\write\@auxout{\string
      \newlabel{#1}{{\@currentlabel}{\thepage}}}}}\@gtempa
   \if@nobreak \ifvmode\nobreak\fi\fi\fi\@esphack}
        \gdef\@eqnlabel{#1}}
\def\@eqnlabel{}
\def\@vacuum{}
\def\draftmarginnote#1{\marginpar{\raggedright\scriptsize\tt#1}}

\def\draft{
        \def\@oddfoot{\sl preliminary draft \hfil
        \rm\thepage\hfil\sl\today\quad\militarytime}
        \let\@evenfoot\@oddfoot \overfullrule 3pt
        \let\label=\draftlabel
        \let\marginnote=\draftmarginnote
   \def\@eqnnum{(\theequation)\rlap{\kern\marginparsep\tt\@eqnlabel}%
\global\let\@eqnlabel\@vacuum}  }

\def\preprint{\twocolumn\sloppy\flushbottom\parindent 2em
        \leftmargini 2em\leftmarginv .5em\leftmarginvi .5em
        \oddsidemargin -.5in    \evensidemargin -.5in
        \columnsep .4in \footheight 0pt
        \textwidth 10.in        \topmargin  -.4in
        \headheight 12pt \topskip .4in
        \textheight 6.9in \footskip 0pt
        \def\@oddhead{\thepage\hfil\addtocounter{page}{1}\thepage}
        \let\@evenhead\@oddhead \def\@oddfoot{} \def\@evenfoot{} }

\def\numberbysection{\@addtoreset{equation}{section}
        \def\theequation{\thesection.\arabic{equation}}}

\def\underline#1{\relax\ifmmode\@@underline#1\else
        $\@@underline{\hbox{#1}}$\relax\fi}

\def\titlepage{\@restonecolfalse\if@twocolumn\@restonecoltrue\onecolumn
     \else \newpage \fi \thispagestyle{empty}\c@page\z@
        \def\thefootnote{\fnsymbol{footnote}} }

\def\endtitlepage{\if@restonecol\twocolumn \else \newpage \fi
        \def\thefootnote{\arabic{footnote}}
        \setcounter{footnote}{0}}  

\catcode`@=12
\relax

\def\figcap{\section*{Figure Captions\markboth
        {FIGURECAPTIONS}{FIGURECAPTIONS}}\list
        {Figure \arabic{enumi}:\hfill}{\settowidth\labelwidth{Figure
999:}
        \leftmargin\labelwidth
        \advance\leftmargin\labelsep\usecounter{enumi}}}
 \relax
\def\tablecap{\section*{Table Captions\markboth
        {TABLECAPTIONS}{TABLECAPTIONS}}\list
        {Table \arabic{enumi}:\hfill}{\settowidth\labelwidth{Table
999:}
        \leftmargin\labelwidth
        \advance\leftmargin\labelsep\usecounter{enumi}}}
 \relax
\def\reflist{\section*{References\markboth
        {REFLIST}{REFLIST}}\list
        {[\arabic{enumi}]\hfill}{\settowidth\labelwidth{[999]}
        \leftmargin\labelwidth
        \advance\leftmargin\labelsep\usecounter{enumi}}}
 \relax
\makeatletter
\newcounter{pubctr}
\def\publist{\@ifnextchar[{\@publist}{\@@publist}}
\def\@publist[#1]{\list
        {[\arabic{pubctr}]\hfill}{\settowidth\labelwidth{[999]}
        \leftmargin\labelwidth
        \advance\leftmargin\labelsep
        \@nmbrlisttrue\def\@listctr{pubctr}
        \setcounter{pubctr}{#1}\addtocounter{pubctr}{-1}}}
\def\@@publist{\list
        {[\arabic{pubctr}]\hfill}{\settowidth\labelwidth{[999]}
        \leftmargin\labelwidth
        \advance\leftmargin\labelsep
        \@nmbrlisttrue\def\@listctr{pubctr}}}
 \relax
\makeatother
\newskip\humongous \humongous=0pt plus 1000pt minus 1000pt

\newif\ifdtup

\relax

\def\be{\begin{equation}}
\def\ee{\end{equation}}
\def\ba{\begin{eqnarray}}
\def\ea{\end{eqnarray}}

\def\a{\alpha}

\def\b{\beta}

\def\g{\gamma}
\def\G{\Gamma}

\def\P{\Pi}

\def\th{\theta}
\def\Th{\Theta}
\def\m{\mu}
\def\n{\nu}

\def\Om{\Omega}
\def\l{\lambda}

\def\s{\sigma}
\def\S{\Sigma}

\def\cN{{\cal N}}

\def\no{\noindent}

\def\qq{\qquad}

\def\IR{\relax{\rm I\kern-.18em R}}

\def \ha {{1\over 2}}

\def \ov {\over}

\def\const{{\rm const.}}

\def\II{\relax{\rm 1\kern-.35em1}}
\def\IR{\relax{\rm I\kern-.18em R}}
\def\inv{^{\raise.15ex\hbox{${\scriptscriptstyle -}$}\kern-.05em 1}}

\def\cF{{\cal F}}

\def\tL{{\tilde L}}

\begin{document}


\renewcommand{\theequation}{\thesection.\arabic{equation}}
\csname @addtoreset\endcsname{equation}{section}

\newcommand{\beq}{\begin{equation}}
\newcommand{\eeq}[1]{\label{#1}\end{equation}}
\newcommand{\ber}{\begin{eqnarray}}
\newcommand{\eer}[1]{\label{#1}\end{eqnarray}}
\newcommand{\eqn}[1]{(\ref{#1})}
\begin{titlepage}
\begin{center}

\hfill hep--th/0609079\\

\vskip .5in

{\large \bf On the velocity and chemical-potential dependence of the
heavy-quark interaction in $\cN=4$ SYM plasmas}

\vskip 0.5in

{\bf Spyros D. Avramis$^{1,2}$},\phantom{x} {\bf Konstadinos
Sfetsos}$^2$\phantom{x} and\phantom{x} {\bf Dimitrios Zoakos}$^{2,*}$
\vskip 0.1in

${}^1\!$
Department of Physics, National Technical University of Athens,\\
15773, Athens, Greece\\

\vskip .1in

${}^2\!$
Department of Engineering Sciences, University of Patras,\\
26110 Patras, Greece\\

\vskip .15in

{\footnotesize {\tt avramis@mail.cern.ch}, \ \ {\tt sfetsos@upatras.gr},
\ \ {\tt dzoakos@upatras.gr}}\\

\end{center}

\vskip .4in

\centerline{\bf Abstract}

\no
We consider the interaction of a heavy quark-antiquark pair
moving in $\cN=4$ SYM plasma in the presence of non-vanishing
chemical potentials. Of particular importance is the maximal length
beyond which the interaction is practically turned off. We propose
a simple phenomenological law that takes into account the velocity
dependence of this screening length beyond the leading order and
in addition its dependence on the R-charge. Our proposal is based
on studies using rotating D3-branes.

\vfill
\no
\hrule width 6.7cm \vskip.1mm{\small \small \small \noindent
$^\ast$\! Currently serving in the Hellenic Armed Forces.}

\end{titlepage}
\vfill
\eject

\def\baselinestretch{1.2}
\baselineskip 10 pt
\noindent

\def\tT{{\tilde T}}
\def\tg{{\tilde g}}
\def\tL{{\tilde L}}


\tableofcontents

\def\baselinestretch{1.2}
\baselineskip 20 pt
\no

\section{Introduction and summary}

Since the early days of the AdS/CFT correspondence \cite{adscft},
there has been considerable interest in applying string-theoretical techniques
for the computation of the interaction
potential of an external
heavy $q\bar{q}$ pair, as given by the expectation value of a
Wilson loop, first demonstrated in \cite{wilsonloop} for the
conformal case. In extensions of this work to non-conformal cases,
it was often found that there exists a maximal length beyond which
the interaction is practically turned off, the archetypal example
being finite-temperature $\cN=4$ SYM theory, described by
non-extremal D3-branes \cite{wilsonloopTemp}. This phenomenon of
complete screening persists \cite{BS} in the $\cN=4$ theory with chemical
potentials, modelled using rotating D3-brane solutions, as well as
in the Coulomb branch of the theory, modelled by continuous
distributions of D3-branes, even though in that latter
case half of the maximal supersymmetries are preserved. Although
such results would be hard to obtain using conventional
field theoretical methods,
their relevance for real QCD was obscured both by the major
qualitative differences between QCD and SYM theories and by the
fact that the quark-gluon plasma (QGP) used for the experimental
study of thermal QCD was until recently thought to be a weakly-coupled collection
of quasiparticles \cite{wqgp}.

This situation changed drastically after the realization that the
QGP is a state of matter best described as a strongly-coupled
liquid, but still exhibiting partial deconfinement of color charges (see e.g. \cite{sqgp} for reviews).
These features motivate the study of various quantities of physical interest within
the gravity/gauge theory correspondence. The best-studied examples
encouraging this line of thought are the calculation of the shear
viscosity in $\cN=4$ SYM \cite{starinets} and the various
calculations of the rate of energy loss of partons moving through
the plasma \cite{drag}-\cite{LinMa}, both of which lead to
results that are in good qualitative agreement with experiment.

This renewed interest shed light into some, often overlooked,
aspects of Wilson-loop calculations of heavy-quark potentials. One
such aspect is that, in real experimental situations, neutral
$q\bar q$ pairs are produced with some velocity relative to the
plasma, a fact that affects the interaction potential of the pair.
Such considerations apply, for example, to the effect of charmonium suppression,
the dissociation of $J / \psi$ mesons due to color screening by the
deconfining medium \cite{matsui,karmehrsatz}, an effect that is
expected to be enhanced when the $J / \psi$
has a nonzero velocity with respect to the plasma.
Although dynamical processes of this kind
would be hard to study using conventional methods,
they can be addressed rather easily within
the gauge/gravity correspondence. In particular, the
non-extremal brane metrics serving as gravity duals
of finite-temperature gauge theories are,
unlike their extremal counterparts, sensitive to boosts in
the directions along the brane where the gauge theory resides.
Then, the computation of the potential interaction of a $q\bar{q}$ pair moving with
respect to the thermal plasma is simply accomplished by applying
the standard AdS/CFT calculation of Wilson loops to such boosted
backgrounds.\footnote{To appreciate this, note, for instance, that a computation of
the potential for a heavy $q\bar q$ pair moving in a hot plasma has been done only for the
case of an abelian gauge theory \cite{chumatsui}.}
The first examples of finite-temperature computations of this kind were
done for a confining non-supersymmetric theory in \cite{peeters} and for $\cN=4$ SYM at vanishing
R-charge in \cite{LRW2,guijosa} (see also \cite{argyres,caceres} for further works).
The results of \cite{LRW2}, in particular, suggest a
universal law
for the behavior of the maximum screening length was suggested,
essentially stating that it depends mainly as $(1-v^2)^{1/4}$ on
the pair's velocity, perfectly valid for ultrahigh velocities.
Given the importance of such a law for building up
phenomenological models for certain effects occurring in hot
plasmas, it is interesting to ask how this law is affected when chemical
potentials are turned on and what are the details of the velocity
dependence. In this paper, we examine precisely these two issues,
within the AdS/CFT correspondence, using rotating D3-brane solutions.
In principle, the maximal screening length can be a function of the plasma
parameters, namely the temperature and the R-charge density and in
addition it depends on the velocity vector of the $q\bar q$ pair.

Our studies suggest a law that can be used in phenomenological models, namely
that the maximal length has the following form in the rest frame of the $q\bar q$ pair
\be
q\bar q{\rm \ rest\ frame}:\phantom{xxx}
L_{\rm max} = \cF(v,\xi) \ {G(\xi) \ov \pi T}\ {1\ov
\g^{{1/2}}}\left(1+ c_1 {\l^2\ov \g} + {c_2 + c_3 \l^4\ov \g
^2}\right)\ ,
\label{jhf9}
\ee
with the definitions
\be
\g = {1\ov \sqrt{1-v^2}}\ ,\qq \xi \sim {J\ov N^2 T^3}\ .
\ee
This form is not very sensitive to the orientation of the axis of the pair as compared
with its velocity. In particular, the average value of the maximal length over
the corresponding angle is, under reasonable assumptions on
the angle distribution of the produced
$q\bar q$ pairs, basically of the same form as well.
In the above, $N$ is the number of colors, $T$ is the plasma
temperature and $J$ is the R-charge density. Also, $\xi$ is the
dimensionless R-charge parameter (the overall numerical value in
its definition will be fixed later), while $\l=\l(\xi)$ is a
dimensionless function of $\xi$ ranging between zero and an upper
bound of order one. Finally, $c_i$, with $i=1,2,3$, are numerical
parameters which, although in our models they have
precisely determined values, relatively small compared to unity,
in more realistic experimental situations they should be
fitted to data. The leading term in Eq. (\ref{jhf9}), which
dominates in the ultrarelativistic (large-$\g$) limit, was already
suggested by the computation of \cite{LRW2}.
However, we shall see
that in our models it makes sense to include the other two terms
since they become important for not too extremely relativistic velocities,
i.e. for $v\lesssim 0.8$. The
function $\cF(v,\xi)$ has a very weak functional dependence and
can be practically taken to be constant. The function
$G(\xi)$ has a behavior of the form $G(\xi)=1+{\cal O}(\xi^2)$
for small $\xi$-values, but can otherwise have either weak or strong dependence
on the parameter $\xi$, depending on the model.
In particular, in cases with $\xi\gg 1$, we have the behavior
$G(\xi)\sim \xi^{-1/3}$, signifying that the screening length
becomes temperature-independent and its scale is set by the R-charge density $J$.
In addition, a dependence proportional to $N^{2/3}$ arises.
For
experimentally-motivated studies however, one may compare ratios of
maximal screening lengths at different velocities so that any
strong dependence on the chemical potential cancels out.
We note that in \cite{caceres} it has been shown that the leading velocity
dependence of the screening length is of the form $(1-v^2)^\nu$,
where the exponent $\nu={1 \ov 4}$ in $\cN=4$ theories,
but is lower than that in certain cases with less supersymmetry.
Based on that we may further generalize \eqn{jhf9} and replace in the exponent
of the leading order term the $1/2$ by the parameter $2 \n$.
We do not believe that this generalization will affect the form
of  the correction terms, as they appear essentially from the analyticity
properties of integrals below.
The same replacement can be done for
the analogous formulae below.

\no
In the rest frame of the plasma,
we should also take into account the Lorentz contraction
of the component of the length along the direction of the motion.
In this case the result is sensitive to the actual distribution probability for the angle
between the $q\bar q$-pair axis and its velocity vector.
The most reasonable assumption is that, that this probability is uniform in the plasma rest
frame. Then, based on our computations, we suggest the following law
\be
{\rm Plasma\ rest \ frame}(1):\phantom{xx}
L_{\rm max} = \cF(v,\xi) \ {G(\xi) \ov \pi T}\ {1+ a \ln \g\ov
\g^{{3/2}}}\left(1+ d_1 {\l^2\ov \g} + {d_2 + d_3 \l^4 \ov \g
^2}\right)\ ,
\label{jhf9abnn}
\ee
where again we have averaged over the angle formed by the $q\bar{q}$-pair's axis
and its velocity vector. The numerical parameters $a$ and $d_i$, $i=1,2,3$
are to be fitted to
possible experimental results. Note that, compared to \eqn{jhf9},
we have a suppression factor $1/\g$ as
as well as the presence of a $\ln \g$ term in the overall coefficient.
As we will show, these are due to the dominance, in taking the average,
of pairs with their axis almost parallel to their velocity vector.

\no
A second possibility, less likely in our opinion, is that the
angle distribution be uniform in the pair's
rest frame. In this case, instead of \eqn{jhf9abnn} we have
\be
{\rm Plasma\ rest \ frame}(2):\phantom{xx}
L_{\rm max} = \cF(v,\xi) \ {G(\xi) \ov \pi T}\ {1\ov
\g^{{1/2}}}\left(1+ d_1 {\l^2\ov \g} + {d_2 + d_3 \l^4 + d_4 \ln \g\ov \g
^2}\right)\ ,
\label{jhf9lab}
\ee
where again we have averaged over the angle. Note that in this case we have no extra
suppression in the length compared to \eqn{jhf9}, with the exception of the appearance of
$\ln \g$ in a subleading term. Hence in this case we have a dominance of angles
near $\Th=\pi/2$, at which the length suffers no Lorentz contraction.

\no
The organization of this paper is as follows. In section 2 we
set up the general formalism for computing the interaction
potential of moving $q\bar q$ pairs in general backgrounds. In
section 3 we review the zero R-charge case and in section 4 we
compute the maximal screening length for more general situations
having non-zero R-charge using rotating D3-brane solutions.
Finally, in section 5 we consider issues concerning the screening
length in the plasma rest frame versus the quark pair rest frame
as well as their average values.

\section{Moving $q\bar{q}$ pairs in the $\cN=4$ plasma}

In the framework of AdS/CFT, Wilson-loop calculations for a
$q\bar{q}$ pair moving with velocity $v$ relative to the plasma
can be carried our either in the rest frame of the plasma where
one considers a string whose endpoints move with velocity $v$ or,
equivalently, in the rest frame of the
$q\bar{q}$ pair, where one considers a string with fixed endpoints
in a boosted background. In what follows, we will employ the second approach.

To describe the method, we begin with a general ten-dimensional
metric of the form
\be
\label{3-1} ds^2 = G_{tt} dt^2 + G_{xx}dx^2 + G_{yy} dy^2 + G_{rr}
dr^2 + \cdots\ ,
\ee
where the metric components do not depend on $t,x$ and $y$ and
where the ellipses denote the remaining terms involving all other
variables. These could be either cyclic variables similar to $x$ and $y$
or angular variables that can be set to some
constant values consistent with their equations of motion.
Orienting the axes so that the $q\bar{q}$ pair moves along the negative $x$
direction, and applying a boost in this direction, we obtain the
metric
\be
\label{3-2}
d\hat{s}^2 = \hat{G}_{tt} dt^2 + \hat{G}_{xx}dx^2 + 2 \hat{G}_{tx} dt dx
+ G_{yy} dy^2 + G_{rr} dr^2 + \cdots\ ,
\ee
where
\be
\label{3-3}
\hat{G}_{tt} = \g^2 (G_{tt} + v^2 G_{xx})\ , \qq \hat{G}_{xx} = \g^2 ( v^2 G_{tt} + G_{xx} )\ , \qq
\hat{G}_{tx} = - \g^2 v ( G_{tt} + G_{xx} )\ .
\ee
The separation length and energy of a
$q\bar{q}$ pair of arbitrary orientation with respect to the
plasma velocity is found by considering a Wilson loop on a
rectangular contour $C$ with one side of length $T$ along the $t$
direction and one side of length $L$ along a spatial direction,
taken for definiteness to be parallel to the $xy$ plane and tilted
at an angle $\Theta$ relative to the $x$ axis, as shown in Fig. 1.
\begin{figure}[t]
\label{fig1}
\begin{center}
\includegraphics [height=8cm]{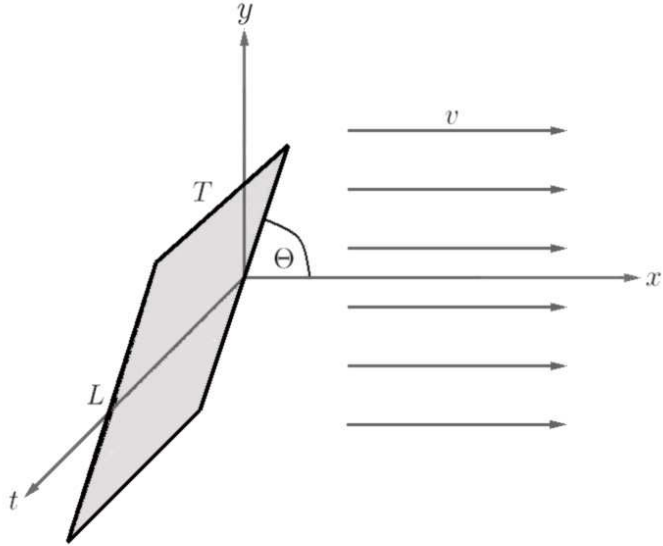}
\end{center}
\vskip -.5 cm \caption{Wilson loop for the energy computation of a
$q \bar{q}$ pair in a moving plasma (shaded rectangle).}
\end{figure}
Then, according to the gauge/gravity correspondence, the
interaction potential energy $E$ of the $q\bar{q}$ pair, as
measured by the Wilson loop expectation $\langle W(C) \rangle$, is
given by \cite{wilsonloop}
\be
\label{3-4}
e^{-i E T} = \langle W(C) \rangle = e^{i S[C]}\ ,
\ee
where
\be
\label{3-5} S[C] = - {1 \ov 2 \pi} \int d \tau d \sigma \sqrt{-
\det g_{\a \b} }\ ,\qq g_{\a\b} = \hat{G}_{\mu\nu}
\partial_\alpha x^\mu \partial_\b x^\nu \ ,
\ee
is the Nambu--Goto action for a string propagating in the boosted
supergravity background, whose endpoints trace the contour $C$. We
note that we will use Minkowski signature throughout the paper.
Fixing reparametrization invariance by taking $(\tau,\s) = (t,y)$,
assuming translational invariance along the $t$ direction and
setting all angles to constants (to specific values consistent
with the equations of motion), the embedding of the string is
described by the functions
\be
\label{3-6}
r=u(y)\ ,\qq x=x(y)\ ,\qq  {\rm rest}   = \const\ ,
\ee
and the boundary condition for the string endpoints
becomes\footnote{Strictly speaking, the above embedding is not
valid at $\Theta=0$, in which case one has to interchange the
roles of $x$ and $y$ i.e. take $(\tau,\s) = (t,x)$ and set
$r=u(x)$ and $y=y(x)$. However, the final formulas for that case,
as quoted after Eq. (\ref{3-14}), can be obtained as limiting
cases with the present embedding.}
\be
\label{3-7}
u \left( \pm {L \ov 2} \sin \Theta \right) = \infty\ ,
\qq x \left( \pm {L \ov 2} \sin \Theta \right) = \pm {L \ov 2}
\cos \Theta\ .
\ee
The Nambu--Goto action is then written in the form
\be
\label{3-8}
S[C] = - {T \ov 2 \pi} \int d y \sqrt{ f_y(u)
+ f_x(u) x^{\prime 2}  + g(u) u^{\prime 2}  }\ ,
\ee
where the prime denotes a derivative with respect to $y$
and\footnote{Note in passing that,
in the limit $v\to 1$ ($\g\to \infty$)
we get the action used in \cite{asjq}
in supergravity computations of the jet quenching parameter.
In taking this limit the extra $\g$ factor is absorbed in
the time dilation of the long side of the Wilson loop.
}
\ba
\label{3-9}
&&f_y(u) = -  \g^2 G_{yy} ( G_{tt} + v^2 G_{xx} )\ , \qq f_x(u) = - G_{tt} G_{xx}\ ,
\nonumber\\
&& g(u) = - \g^2 G_{rr} ( G_{tt} + v^2 G_{xx} )\ .
\ea
Since actions of the precise form \eqn{3-8} have been encountered
in the static Wilson-loop calculations for $x'=0$ in  \cite{BS} (see section 1)
and \cite{HSZ} (see section 5, where $x$ played the role of an angular variable),
we briefly review the results without entering
into the intermediate computational details. Independence of the
Lagrangian from $y$ and $x$ leads to two first-order equations
expressing the conservation of the ``energy'' and the
momentum $\pi_x$ conjugate to $x(y)$. Integrating these equations,
one obtains \cite{HSZ}
\ba
\label{3-10}
L \sin \Theta &=& 2 \sqrt{1-\pi_x^2}\ f_y^{1/2} (u_0)
\int_{u_0}^\infty { du\ov f_y(u)} \sqrt{g(u) \ov  F(u) }\ ,
\nonumber\\
L \cos \Theta &=& 2 \pi_x\ f_x^{1/2}(u_0) \int_{u_0}^\infty
{du \ov f_x(u)} \sqrt{g(u)  \ov F(u)}\ ,
\ea
where the integration constant $u_0$ is the value where $u(y)$ develops a minimum.
Substituting them in the Nambu--Goto action, one
finds the $q\bar{q}$ potential energy
\ba
\label{3-11}
E = {1 \ov \pi} \int^\infty_{u_0} du \sqrt{g(u) \ov F(u)} - E_0\ .
\ea
where $E_0$ is the subtraction term corresponding to the energy of two disconnected
worldsheets, to be discussed below. In the above, the function $F(u)$
is defined by\footnote{Compared to \cite{HSZ}, this function has been
rescaled by a factor of $f_y(u)$.}
\ba
\label{3-12}
F(u) =  1 - \pi_x^2 {f_x(u_0) \ov f_x(u)} - (1-\pi_x^2) {f_y(u_0)\ov f_y(u)}\ .
\ea
and satisfies $F(u_0)=0$. The system of Eqs. \eqn{3-10}
implicitly determines the parameters $u_0$ and $\pi_x$ in terms
of the separation length $L$ and the angle $\Th$, the latter clearly
only via the combinations $\cos^2\Th$ and $\sin^2\Th$.
Substituting the resulting values into \eqn{3-11} gives the
potential energy for the same parameters.
Note that Eqs. \eqn{3-10}-\eqn{3-12}
are invariant under the symmetry operation
$\Th\to {\pi\ov 2}-\Th$ followed by the exchange of the $x$ and
$y$ indices and the renaming
$\pi_x^2\to 1-\pi_x^2$.

\no
To avoid unnecessary complications, we will restrict to the cases
where the spatial side of the Wilson loop is either perpendicular
($\Theta={\pi \ov 2}$) or parallel ($\Theta=0$) to the plasma
velocity and only discuss the general case in section 5.
In the first case, we have $\pi_x=0$ and so the length
and energy read \cite{BS}
\be
\label{3-13}
L = 2 f_y^{1/2} (u_0) \int_{u_0}^\infty du \sqrt{g(u) \ov f_y(u) [ f_y(u) - f_y(u_0) ] }\ ,
\ee
and
\be
\label{3-14}
E = {1 \ov \pi} \int^\infty_{u_0} du \sqrt{g(u) f_y(u) \ov f_y(u) - f_y(u_0)} - E_0\ ,
\ee
while in the second case we have $\pi_x=1$ and the length and
energy are given by the above formulas with the replacement
$f_y \to f_x$.

\no
At this point, some clarifying comments are in order. A first
issue is how much the string may stretch into the radial
direction, i.e. which is the minimum allowed value for the turning
point $u_0$. To examine this issue, we note that in the geometries
under consideration there appear several special values of the
radius, namely (i) the horizon radius $u_H$ where $g(u)$ blows up,
(ii) the radius $u_e$ where $f_x(u)$ becomes zero, which in our examples,
depending on the angle,
coincides with the maximum or minimum values of the ergosphere
(which, in the absence of rotation, coincides with $u_H$) and
(iii) the velocity-dependent radius $u_v$ where $g(u)$ and
$f_y(u)$ become zero\footnote{The significance of the radius $u_v$
has been pointed out in \cite{drag} for one-quark configurations
and in \cite{LRW2,guijosa,argyres} for two-quark configurations.
Roughly, $u=u_v$ acts as a hypersurface for the boosted metric,
beyond which the string cannot penetrate unless a special choice
of integration constant is employed in the one-quark case
\cite{drag}.} (which, in the zero-velocity limit, coincides with
$u_H$). In our examples, these radii always satisfy $u_H \leqslant
u_e \leqslant u_v$. While it is quite obvious that taking $u_0 \ge
u_v$ is always consistent, some care is needed in order to examine
whether we can take $u_0 < u_v$. To this end, we distinguish
between two cases. When $\Th \ne 0$, the first of Eqs. \eqn{3-10}
is always relevant and the factor $\sqrt{f_y(u_0)}$ in that
equation becomes imaginary for $u_0 < u_v$, therefore excluding
that possibility. On the other hand, when $\Th = 0$, the first of
Eqs. \eqn{3-10} becomes irrelevant and the behavior of the string
is solely determined by the function $x(u)$ with $x'(u) \sim
\sqrt{f_x(u_0)} {\sqrt{g(u)} \ov f_x(u) \sqrt{F(u)}}$. If we take
$u_0 < u_v$ then the string will first reach the radius $u_v$
where $x'(u)$ becomes zero so that the string develops a cusp and
cannot be further extended towards $u_0$. If one wishes to include
such configurations, one must thus cut off the region $u_0
\leqslant u < u_v$ by replacing the lower limit of the
integrations from $u_0$ to $u_v$ for all $u_0<u_v$. However, since
(i) string configurations with cusps raise questions about the
validity of the Nambu--Goto approach since the string curvature
diverges there, (ii) the maximal value of $L(u_0)$, which is the
main quantity of interest in this paper, corresponds to a value
$u^{\rm c}_0$ of $u_0$ that is always larger than $u_v$, and (iii)
the region $u_0<u^{\rm c}_0$ where cusps may appear corresponds to
a branch of the solution (with the same prescribed boundary
conditions \eqn{3-7}) which is energetically unfavorable and thus
presumably unstable, we will altogether ignore the presence of
cusps as they are irrelevant for our screening problem.

\no
Another issue refers to the choice of subtraction term $E_0$ in
\eqn{3-11}. In the zero-velocity case, the subtraction term is
unique and represents two straight strings stretching from
infinity to the horizon. In the case of nonzero velocity, however,
the above configuration is no longer a solution to the equations
of motionand one must find an appropriate subtraction term that
(i) removes the divergence, (ii) is consistent with the equations
of motion, (iii) is independent of the separation length and (iv)
reduces in the zero-velocity limit to two straight strings
stretching to the horizon. These criteria are satisfied by any
configuration of two disconnected strings interpolating between
the drag configuration of \cite{drag} (and its generalizations
\cite{moredrag} for the case of rotating branes), which
represents a curved string stretching up to the horizon $u_H$ and
a configuration of one straight string stretching up to the
radius $u_v\geqslant u_H$. Note that in order for the strings to end at a
radius larger than $u_H$, one needs to assume the presence of a
second flavor brane at that radius, in which case each string
should be regarded as dual to a meson composed out of a heavy and
a light quark (like a bottom and an up or a down quark).
Be that as it may, there seems to exist an infinite
family of subtraction terms \cite{LRW3} consistent with the above
requirements. The maximum subtraction energy \cite{guijosa}
corresponds to the drag configuration and is given by
\be
\label{reg1}
E_0 = {1\ov \pi} \int_{u_H}^{\infty} du\ ,
\ee
while
the minimum subtraction energy corresponds to the straight-string
configuration and reads
\be
\label{reg2} E_0 = {1\ov \pi}
\int_{u_v}^{\infty} du \sqrt{g(u)}\ .
\ee
Among the possible
subtraction terms, the one corresponding to the drag configuration
seems to be the most sensible one, as it is the only solution available in the
absence of an extra flavor brane.

\section{Screening in the zero--R-charge case}

In order to establish our notation and present
some extra details, we first review  the zero--R-charge case, examined
in \cite{LRW2} and also in \cite{guijosa}. In this case, the
background geometry has metric
\be
 ds^2 = {r^2\ov R^2} \left[-\left(1-{\m^4  \ov r^4}\right) dt^2 + dx^2 + dy^2 + dz^2\right]
+ {R^2 r^2\ov r^4-\m^4}\ dr^2 + R^2d\Om_5^2\ ,
\label{coosnt}
\ee
where the horizon is located at $r=\m$ and the Hawking temperature
is $T={\m\ov \pi R^2}$. Then we have for the various functions
\be
f_y(u) = {1\ov R^4}(u^4 - \g^2 \m^4)\ , \qq f_x(u) ={1\ov R^4}( u^4 - \m^4)\ ,
\qq g(u) = {u^4-\g^2 \m^4 \ov u^4-\m^4}\ .
\label{hcd1}
\ee

\subsection{Motion perpendicular to the $q\bar q$-pair axis}

Let us first consider the $\Theta={\pi \ov 2}$ case corresponding
to plasma velocities perpendicular to the axis of the $q\bar q$
pair. Substituting $f_y(u)$ and $g(u)$ from \eqn{hcd1} into Eqs.
\eqn{3-13} and \eqn{3-14}, introducing the dimensionless length
and energy parameters
\be
\label{dimensionless}
\ell = \pi T L \ ,\qq  \varepsilon = {1 \ov R^2 T} E \ ,
\ee
and trading $u$ and $u_0$ for the dimensionless variables $y={u
\ov \m}$ and $y_0={u_0 \ov \m} \geqslant \sqrt{\g}$, we obtain for the length
\ba
\ell(v,y_0) &=& 2 \sqrt{y_0^4 - \g^2}
\int_{y_0}^\infty {d y \ov \sqrt{ ( y^4 - 1) (y^4 - y_0^4)}}
\nonumber\\
&=& {2 \sqrt{2} \pi^{3/2} \ov \Gamma(1/4)^2}\ {\sqrt{y_0^4-\g^2} \ov y_0^3}\
{}_2F\!_1 \left({1 \ov 2},{3 \ov 4},{5 \ov 4},{1 \ov y_0^4} \right)\ ,
\label{jdj1}
\ea
where ${}_2F\!_1(a,b,c,x)$ denotes the standard hypergeometric function. For the energy,
use of the maximum subtraction \eqn{reg1} leads to the result
\ba
\varepsilon(v,y_0) &=& \int_{y_0}^\infty dy \left[ {y^4 - \g^2 \ov \sqrt{(y^4 - 1)(y^4-y_0^4)} }
- 1 \right] - (y_0-1)
\nonumber\\
&=& - {\sqrt{2} \pi^{3/2} \ov \Gamma(1/4)^2} \left[ y_0\
{}_2F\!_1\left(-{1 \ov 4},{1 \ov 2},{1 \ov 4},{1 \ov y_0^4} \right) + {\g^2
\ov y_0^3}\ {}_2F\!_1\left({1 \ov 2},{3 \ov 4},{5 \ov 4},{1 \ov y_0^4}
\right) \right] + 1\ ,
\label{enmax}
\ea
while use of the minimum subtraction \eqn{reg2} yields
\ba
\varepsilon(v,y_0) &=& \int_{y_0}^\infty dy \left[ { y^4 - \g^2 \ov \sqrt{(y^4 - 1)(y^4-y_0^4)} }
- \sqrt{ {y^4 - \g^2 \ov y^4 - 1 } } \right] - \int_{\sqrt{\g}}^{y_0} dy \sqrt{ {y^4 - \g^2 \ov y^4 - 1 } }
\nonumber\\
&=& - {\sqrt{2} \pi^{3/2} \ov \Gamma(1/4)^2} \biggl[ y_0\
{}_2F\!_1\left(-{1 \ov 4},{1 \ov 2},{1 \ov 4},{1 \ov y_0^4} \right) + {\g^2
\ov y_0^3}\ {}_2F\!_1\left({1 \ov 2},{3 \ov 4},{5 \ov 4},{1 \ov y_0^4}
\right)
\label{enmin}\\
&& \qq\qq\quad - 2 \sqrt{\g}\
{}_2F\!_1\left({1 \ov 2},-{1 \ov 4},{5 \ov 4},{1 \ov \g^2} \right) \biggr]\ ,
\nonumber
\ea
Note that, according to the discussion of section 2, no cusps occur and that
the two expressions above differ only by the last term which is
velocity dependent, but totally independent of the parameter $u_0$. As a result the
two expressions differ by an $\ell$-independent term and the force between the
quarks, given by minus the derivative of the energy with respect to the separation length,
is the same no matter which expression we use.
In the static limit in which $\g=1$, using properties of the hypergeometric
functions, the expressions \eqn{enmax} and \eqn{enmin} are found to coincide.
Also for $\ell\to 0$, both go to the conformal result of \cite{wilsonloop}.

\no
Fig. 2(a) shows the dimensionless energy $\varepsilon$ as a
function of the dimensionless separation length $\ell$ for
various values of the velocity $v$. We see that $\varepsilon$ is a
double-valued function of $\ell$ whose upper branch corresponds to
a presumably metastable, at best, configuration and is to be discarded.
As in the static case with $v=0$
\cite{wilsonloopTemp}, there is a maximal length $\ell_{\rm
max}(v)$ beyond which there is complete screening of the potential
since the only allowed configuration is the zero-energy
configuration of two disconnected worldsheets. Its numerical value
for $v=0$ is $\ell_{\rm max}(0)=0.869$ was given in
\cite{BS,LRW2}. For ultrahigh velocities it goes to zero as
$\ell_{\rm max}(v\to 1) = 0.743 \g^{-1/2}$ as first noted in
\cite{LRW2}. There are various ways to see that. In particular,
assuming that $y_0\gg 1$,
from \eqn{jdj1} and the fact that the hypergeometric function
becomes unity at the origin, we see that $\ell \simeq {2\sqrt{2}
\pi^{3/2}\ov \G(1/4)^2} {\sqrt{y_0^4-\g^2}\ov y_0^3}$.
Maximization of this with respect to $y_0$ gives $y_0^2=\sqrt{3}
\g$, which upon substitution gives $\ell_{\rm max}(v\to 1) = {4
\pi^{3/2} \ov 3^{3/4} \G(1/4)^2} \g^{-1/2}\simeq 0.743 \g^{-1/2}$,
as stated above.
\begin{figure}[t]
\label{fig2}
\begin{center}
\begin{tabular}{ccc}
\includegraphics [height=4.8cm]{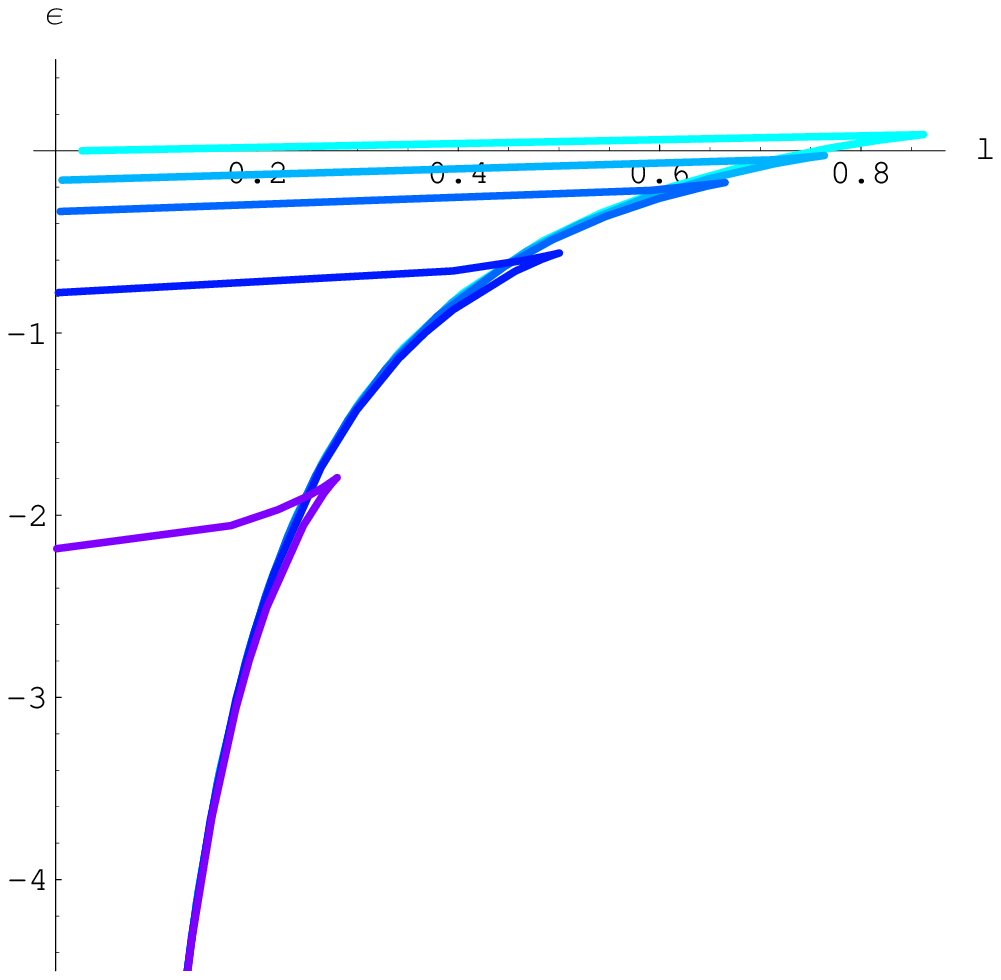}
& \includegraphics[height=4.8cm]{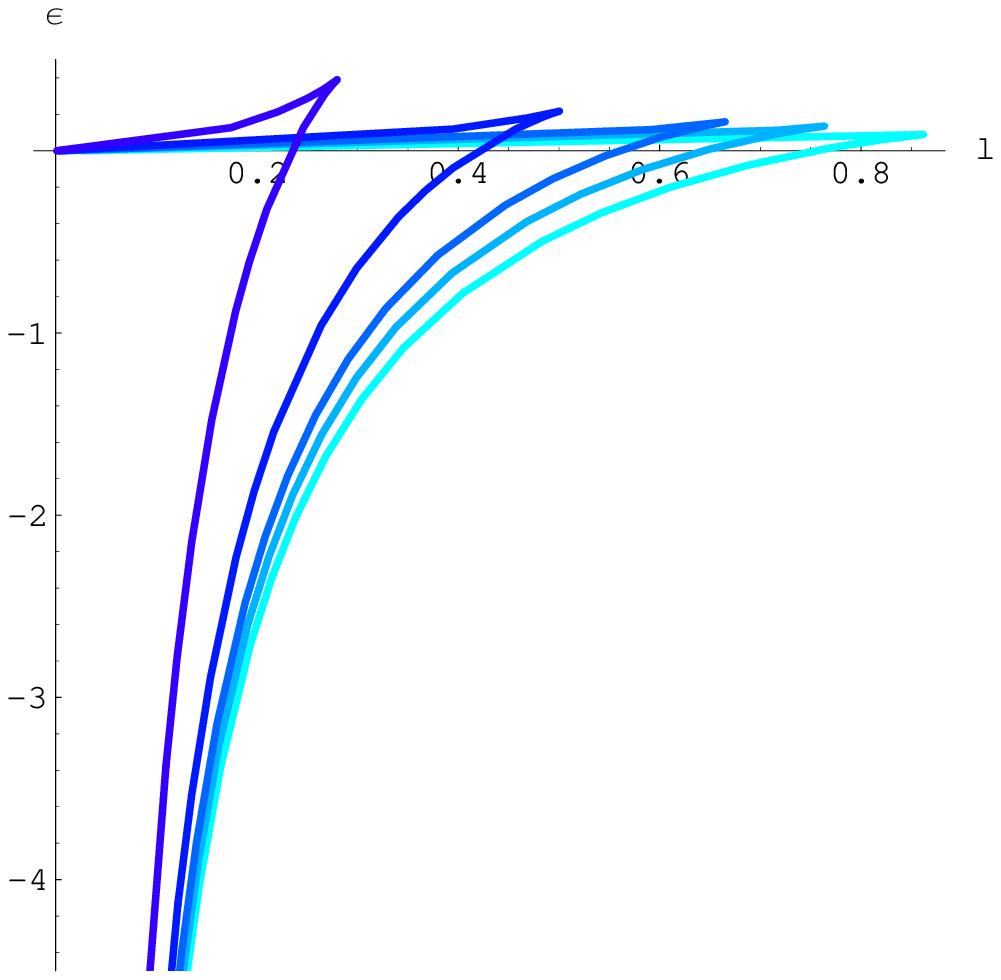}
& \includegraphics[height=4.8cm]{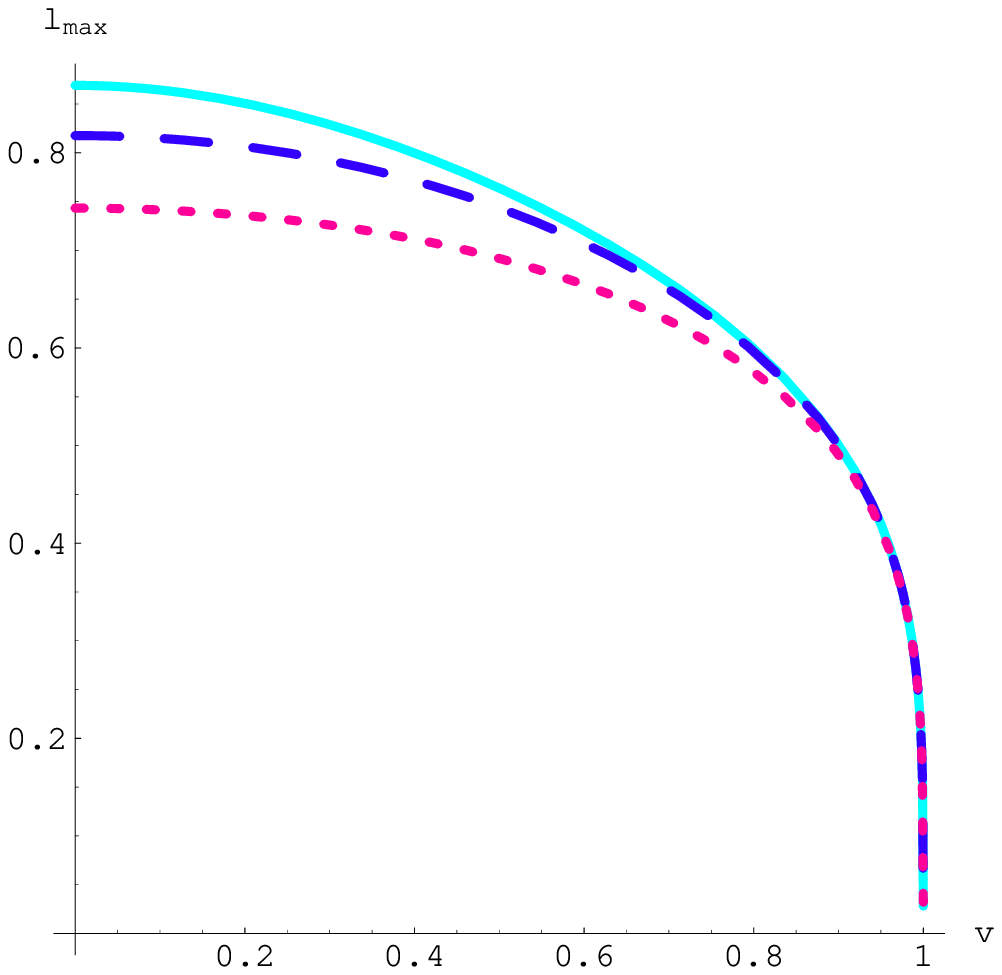}\\
(a) & (b) & (c)
\end{tabular}
\end{center}
\vskip -.5 cm \caption{(a,b) Potential energy $\varepsilon$ plotted
as a function of the separation $\ell$ for the values $v=0$,
$0.5$, $0.7$, $0.9$ and $0.99$ (right to left) using the maximum
and minimum subtraction prescriptions respectively.
(c) Plots of various approximations to $\ell_{\rm
max}(v)$, namely the result of numerical computation (solid), the
approximation \eqn{jhf9} (dashed) and its leading-order form
(dotted).}
\end{figure}

\no
Besides the maximal length, there also exists a value $\ell_{\rm s}(v)$
of the length \cite{LRW2,guijosa}
(its value for $v=0$ is $\ell_{\rm s}(0)=0.754$ \cite{BS}) after
which the lower branch of $\varepsilon$ takes positive values.
The region beyond this point is also to be discarded, since
the free configuration becomes energetically favored there, and hence this
length could be also used as a definition of the screening
length. Using the regularization \eqn{enmax} for the energy there is a velocity
$v_c\simeq 0.447$ for which the maximum length and $\ell_{\rm s}$ coincide \cite{guijosa} and after which
there is no distinction between them.
Then, as shown in Fig. 2(a), for small velocities the length $\ell_{\rm s}(v)$
differs from the maximal one by about ten percent and,
as the velocity is increased, the difference becomes
smaller and vanishes for the above value of the velocity $v_c$.
Since $\ell_{\rm s}$ practically coincides with $\ell_{\max}$ for $v \gtrsim 0.35$,
and since its study necessitates examining both $\varepsilon$ and $\ell$
thus requiring more involved numerics,  in
this paper we will define the screening length as the maximal
length allowed.

\no
The above results for $\ell_{\rm max}(v)$ served as a motivation
for the authors of \cite{LRW2} to suggest that the velocity
dependence of the maximal length is given by the empirical formula
\be
\ell_{\rm max}(v) = \cF(v) (1-v^2)^{1/4}\ ,
\label{ih3}
\ee
where $\cF(v)$ is a function with relatively weak dependence on
$v$, with $\cF(0)=0.869$ and $\cF(1) = 0.743$. With an analysis
that we will perform in a more general setting in the next
section, we will find an improved version of \eqn{ih3}
\be
\ell_{\rm max}(v) = \cF(v) (1-v^2)^{1/4}\left[1+c_2
(1-v^2)^{1/2} \right]\ , \label{ih32}
\ee
which is of the form \eqn{jhf9}. For our model the
numerical coefficient is computed in more general situations below
and takes the value $c_2=0.100$. In that case $\cF(v)$ is indeed a
practically flat function of $v$ since $\cF(0)=0.790$ and $\cF(1)
= 0.743$. Plots of $\ell_{\rm max}(v)$ in the various
approximations considered are shown in Fig. 2(c).

\subsection{Motion parallel to the $q\bar q$-pair axis}

Let us consider next the $\Theta=0$ case, corresponding to plasma
velocities along the axis of the $q\bar q$ pair. Substituting
$f_x(u)$ and $g(u)$ in the corresponding equations and employing
the same rescalings as before, we find for the length
\ba
\ell(v,y_0) & =& 2 \sqrt{y_0^4-1} \int_{y_0}^\infty {d y \ov y^4 -
1} \sqrt{y^4 - \g^2 \ov y^4 - y_0^4}\ ,
\nonumber\\
& = & {2 \sqrt{2} \pi^{3/2}\ov \G(1/4)^2}\ {\sqrt{y_0^4-1}\ov
y_0^3}\ F\!_1\left({3\ov 4},-\ha,1,{5\ov 4},{\g^2\ov y_0^4}, {1\ov
y_0^4}\right)\ ,
\ea
where $F\!_1(a, b_1, b_2, c, x, y)$ is the Appell hypergeometric
function. For the energy, use of the maximum subtraction \eqn{reg1}
and the minimum subtraction \eqn{reg2} leads to
\ba
\varepsilon(v,y_0) & = &
\int_{y_0}^\infty dy \left( \sqrt{{y^4 - \g^2 \ov y^4-y_0^4}} - 1
\right) - (y_0-1)\ ,
\nonumber\\
& = & -{\sqrt{2} \pi^{3/2}\ov \G(1/4)^2} y_0\ {}_2F\!_1\left(-\ha ,
-{1\ov 4},{1\ov 4},{\g^2\ov y_0^4}\right) + 1\ .
\ea
and
\ba
\varepsilon(v,y_0) & = &
\int_{y_0}^\infty dy \left( \sqrt{ y^4 - \g^2 \ov y^4-y_0^4} - \sqrt{ {y^4 - \g^2 \ov y^4 - 1 } }
\right) - \int_{\sqrt{\g}}^{y_0} dy \sqrt{ {y^4 - \g^2 \ov y^4 - 1 } }\ ,
\nonumber\\
& = & -{\sqrt{2} \pi^{3/2}\ov \G(1/4)^2} \left[ y_0\ {}_2F\!_1\left(-\ha ,
-{1\ov 4},{1\ov 4},{\g^2\ov y_0^4}\right) -  2 \sqrt{\g}\
{}_2F\!_1\left(\ha , -{1\ov 4},{5\ov 4},{1\ov \g^2}\right)\right]\ .
\ea
respectively. Note that, according to the discussion of section 2,
one may take $y_0<y_v=\sqrt{\g}$ in which case there may occur
cusps. However, the branch where cusps occur is irrelevant for our problem.
This subtlety aside, the qualitative behavior discussed in the
previous subsection remains virtually
unchanged, with only small deviations from the $\Theta={\pi \ov
2}$ case. For instance, the numerical coefficient for the $v \to
1$ behavior changes to $0.837$ but nevertheless the velocity
dependence remains of the same form. For this reason, in the rest of this
paper we will consider in detail only the $\Theta = {\pi \ov 2}$ case.

\section{Screening in the case of nonzero R-charges}

In light of the above results, a natural question arising
is whether the $(1-v^2)^{1/4}$ velocity scaling law is universal
in more generalized settings, one of which is the $\cN=4$ theory
with non-vanishing R-charges. In what follows, we will briefly
review the relevant supergravity backgrounds and then we will
address the questions (i) whether this velocity dependence
persists in the presence of R-charges and (ii) what is the nature
of the dependence of the maximal length on the R-charge density.

\subsection{Non-extremal rotating branes}

According to AdS/CFT, the gravity duals to $\cN=4$, $SU(N)$ SYM
theory at finite temperature and R-charge are given by
non-extremal rotating D3-brane solutions found in
\cite{rotatingbranesmore,cvetic}. In the conventions of \cite{rs}
which we here follow, the field-theory limit of these solutions is
characterized by the non-extremality parameter $\m$ and the
angular momentum parameters $a_i$, $i=1,2,3$, which are related to
three R-charges (or chemical potentials) in the gauge theory.
Below we review two special cases of these solutions,
corresponding to two equal nonzero angular momenta and one nonzero
angular momentum, for which we state the relations between the
supergravity and gauge-theory parameters as well as the
restrictions on the physical ranges of these parameters as
dictated by thermodynamic stability \cite{rotatingbranesthermo}.
For more details on the calculations, the reader is referred to
\cite{asjq}.

\subsubsection{ Two equal nonzero angular momenta}

We first examine the case of two equal nonzero angular momenta,
which we can take as $a_2=a_3=r_0$.
The metric is given by
\ba
&& ds^2 = H^{-1/2} \left[-\left(1-{\m^4 H \ov R^4}\right) dt^2 + dx^2 + dy^2 + dz^2\right]
+ H^{1/2} {r^4(r^2-r_0^2 \cos^2\th)\ov (r^4-\m^4)(r^2-r_0^2)}\ dr^2
\nonumber\\
&& + H^{1/2}\Big[(r^2-r_0^2\cos^2\th )d\th^2 + r^2 \cos^2\th d\Om_3^2
+ (r^2-r_0^2)\sin^2\th d \phi_1^2
\label{sphe}
\\
&& - \ 2 {\m^2 r_0\ov R^2} \ dt \cos^2\th (\sin^2\psi d \phi_2 + \cos^2\psi d \phi_3)\Big]\ ,
\nonumber
\ea
where
\be
H={R^4\ov r^2 (r^2-r_0^2\cos^2 \th)} \
\ee
and $d\Om^2_3  =d\psi^2 + \sin^2 \psi d \phi_2^2 + \cos^2 \psi d
\phi_3^2$ is the $S^3$ line element. The horizon is located at
\be
r_H = \m\ .
\ee
On the gravity side, the above solution is characterized by the
two parameters $\m$ and $r_0$. On the dual gauge-theory side, it
is most natural to use the Hawking temperature $T$ and the angular
momentum density $J$, which correspond to the temperature and
R-charge density of the gauge theory i.e. to the
canonical-ensemble thermodynamic variables. On both sides, it is
convenient to trade $r_0$ and $J$ for the dimensionless parameters
\be
\l = {r_0 \ov \m}\ ,\qq \xi = {6 \sqrt{3} J \ov \pi N^2 T^3}\ .
\label{rml1}
\ee
The two sets of parameters $(\m,\l)$ and $(T,\xi)$ are related by
\be
\label{rml1a}
\m = \pi R^2 T \sqrt{1+Q^2}\ , \qq \l = {Q \ov \sqrt{1+Q^2}}\ ,
\ee
where $F(\xi)$ is the function
\be
Q(\xi) ={(\xi + \sqrt{1+\xi^2})^{1/3} - (\xi + \sqrt{1+\xi^2})^{-1/3}\ov \sqrt{3}}\ .
\ee
Substituting this into \eqn{rml1a} we obtain the explicit expression for $\l=\l(\xi)$.
The constraints imposed by thermodynamic stability on the
parameters $\l$, $\xi$ and $Q$ as follows
\be
0 \leqslant \l \leqslant 1\ ,\qq 0 \leqslant \xi < \infty\ , \qq 0\leqslant Q < \infty\ .
\ee

\subsubsection{ One nonzero angular momentum}

We next examine the case of only one nonzero angular momentum,
which we can take as $a_1=r_0$. The metric is given by
\ba
&& ds^2 = H^{-1/2} \left[-\left(1-{\m^4 H \ov R^4}\right) dt^2 + dx^2 + dy^2 + dz^2\right]
+ H^{1/2} {r^2(r^2+r_0^2 \cos^2\th)\ov r^4+r_0^2 r^2 -\m^4}\ dr^2
\nonumber\\
&& + H^{1/2}\Big[(r^2+r_0^2\cos^2\th )d\th^2 + r^2 \cos^2\th d\Om_3^2
+ (r^2+r_0^2)\sin^2\th d \phi_1^2
\label{discc}
\\
&& - \ 2 {\m^2 r_0\ov R^2} \sin^2\th dt d\phi_1\Big]\ ,
\nonumber
\ea
where
\be
H={R^4\ov r^2 (r^2+r_0^2\cos^2 \th)} \
\ee
and the $S^3$ line element is as before. The horizon is located at
$r=r_H$ with
\be
\label{2-9}
r_H^2 = {1 \ov 2} \left( - r_0^2 + \sqrt{r_0^4 + 4 \m^4} \right)\ .
\ee
As before, the solution is characterized by $\m$ and $r_0$ on the
gravity side and by $T$ and $J$ on the dual gauge-theory side.
Introducing again dimensionless variables according to
\be
\l = {r_0 \ov \m}\ ,\qq \xi = {2 \sqrt{2} J \ov \pi N^2 T^3}\ ,
\ee
we find that the two sets of parameters are related by
\be
\m = \pi R^2 T  Q^{3/4} (2-Q)^{1/4}\ ,\qq \l = {\sqrt{2} (Q-1)^{1/2} \ov Q^{1/4} (2-Q)^{1/4}}\ ,
\label{rml2}
\ee
where $Q(\xi)$ is the lower branch of the solution to the equation
$Q^4 (Q-1)(Q-2)+\xi^2 = 0$, which admits a perturbative expansion, near $\xi=0$, as
$Q=1+{\cal O}(\xi^2)$.
Again, one may substitute this into \eqn{rml2} to obtain a
perturbative expansion for $\l=\l(\xi)$. Finally, thermodynamic
stability constrains $\l$, $\xi$ and $Q$ as follows
\be
0 \leqslant \l \leqslant \l_0 \simeq 1.46 \ , \qq 0 \leqslant \xi \leqslant \xi_{\rm max} \simeq 1.33\ ,\qq
1 \leqslant Q \leqslant Q_{\rm max} \simeq 1.73 \ .
\ee

\subsection{Velocity and R-charge dependence of the maximal length}

After the above preliminaries, we are ready to examine the problem
of screening in the presence of nonzero R-charges. The new
parameter entering the problem in this case is the supergravity
angular momentum parameter $\l$ or, equivalently, the gauge-theory
R-charge density $\xi$. Therefore, the dimensionless length and
energy have the form $\ell(v,\xi,y_0)$ and
$\varepsilon(v,\xi,y_0)$, while the maximal length is of the form
$\ell_{\rm max}(v,\xi)$. In the rest of this section we focus on
the maximal length and on its dependence on the velocity and
R-charge.

\no
As a general remark we note that the leading high-velocity behavior
of the screening length in the presence of R-charges will be the same as that in their
absence, since it occurs for large values of the parameter $u_0$
proportional to $\g^{1/2}$, for which the geometry behaves as if it was $AdS_5 \times S^5$.
This was also noted in \cite{caceres}
from a five-dimensional perspective\footnote{We emphasize that
results from the Nambu--Goto actions
with a ten-dimensional background metric are generically
different than those obtained with the background
metric replaced by the five-dimensional one corresponding to a lower dimensional
supergravity theory. A recent such example is the computation of the jet quenching
parameter in the presence of R-charges. The results of \cite{asjq,AEM}, where a
ten-dimensional approach was followed agree only qualitatively with those of
\cite{LinMa}, whose approach is a five-dimensional one.}
and is apparent when the comparison is done
at constant energy density $\epsilon={3N^2 \m^4 \ov 8\pi^2 R^8}$ above extremallity,
in which case the (dimensionful) maximal length is of the form
$L_{\max} \sim {(1-v^2)^{1/4} \ov \m} \sim{(1-v^2)^{1/4} \ov \epsilon^{1/4}}$,
with the same coefficient as in the zero R-charge case. However, in this paper we compare
quantities at constant temperature so that
the proportionality factor is apparently different and depends on the R-charges.

\subsubsection{Two equal nonzero angular momenta}

For this case, the functions $f_y(u)$ and $g(u)$ are given by
\be
f_y(u) = {1\ov R^4} \left[ u^2(u^2 - r_0^2 \cos^2 \theta) - \g^2
\m^4 \right]\ , \quad g(u) = {u^2 [u^2(u^2-r_0^2 \cos^2 \theta) -
\g^2 \m^4] \ov (u^4-\m^4)(u^2-r_0^2)}\ ,
\ee
where for consistency with the corresponding equations of motion
the angular variable $\th$ must be either $0$ or $\pi/2$. The
expressions for the dimensionless length $\ell$ are as follows.

\no $\bullet$ For the case with $\th=0$, after a change of
integration variable, the dimensionless separation length takes
the form\footnote{In the limiting case $\l=1$ (corresponding to
$\xi \to \infty$), the expression \eqn{ca3i} and the similar one
\eqn{ca4i} below can be evaluated exactly in terms of complete
elliptic integrals. However, no such exact formulas are available
for the integrals \eqn{cas2} and \eqn{cas1} further below.}
\ba
\ell(v,\xi,y_0)
&\! = \! & G(\xi)\ {\sqrt{y_0^4 - \l^2 y_0^2 - \g^2}\ov y_0^3}
\nonumber\\
&& \phantom{} \times
\int_{1}^\infty {dz\ov \sqrt{ ( z^2 - 1/y_0^4) ( z - \l^2/y_0^2) (z-1) (z + 1 - \l^2/y_0^2)}}\ .
\label{ca3i}
\ea
At this point, we are ready to justify for this model the proposed
empirical formula \eqn{jhf9} for the screening length. Since
maximization of \eqn{ca3i} with respect to $y_0$ cannot be carried
out analytically, we employ a perturbative method that proceeds as
follows. We first assume that the maximum, for large enough $\g$,
is at some value of $y_0$ of leading order $\g^{1/2}$, to be
justified by the result, and then we expand the integrand into
inverse powers of $y_0^2$ up to second order. The resulting
expression can then be analytically maximized with respect to $y_0$ and then
the result for $\ell_{\rm max}$ follows by substitution. We find
an expression of the form \eqn{jhf9} with
\be
c_1 = -0.10525\ ,\qq c_2 = 0.100\ ,\qq c_3= -0.01507 \ ,
\ee
where we preferred to give, to a certain accuracy, the numerical
values of the various coefficients, instead of presenting their
explicit expressions in terms of gamma functions. For $\cF(v,\xi)$
we give its numerical values in the four corners in the
$(v,\xi)$-plane
\be
\cF(0,0)= 1.06\ ,\qq \cF(0,\infty)= 0.96\ ,\qq \cF(1,0)= 1\ ,\qq
\cF(1,\infty)= 1\ ,
\ee
showing that it is indeed a very slowly varying function.
\begin{figure}[t]
\label{fig3}
\begin{center}
\begin{tabular}{ccc}
\includegraphics [height=4.8cm]{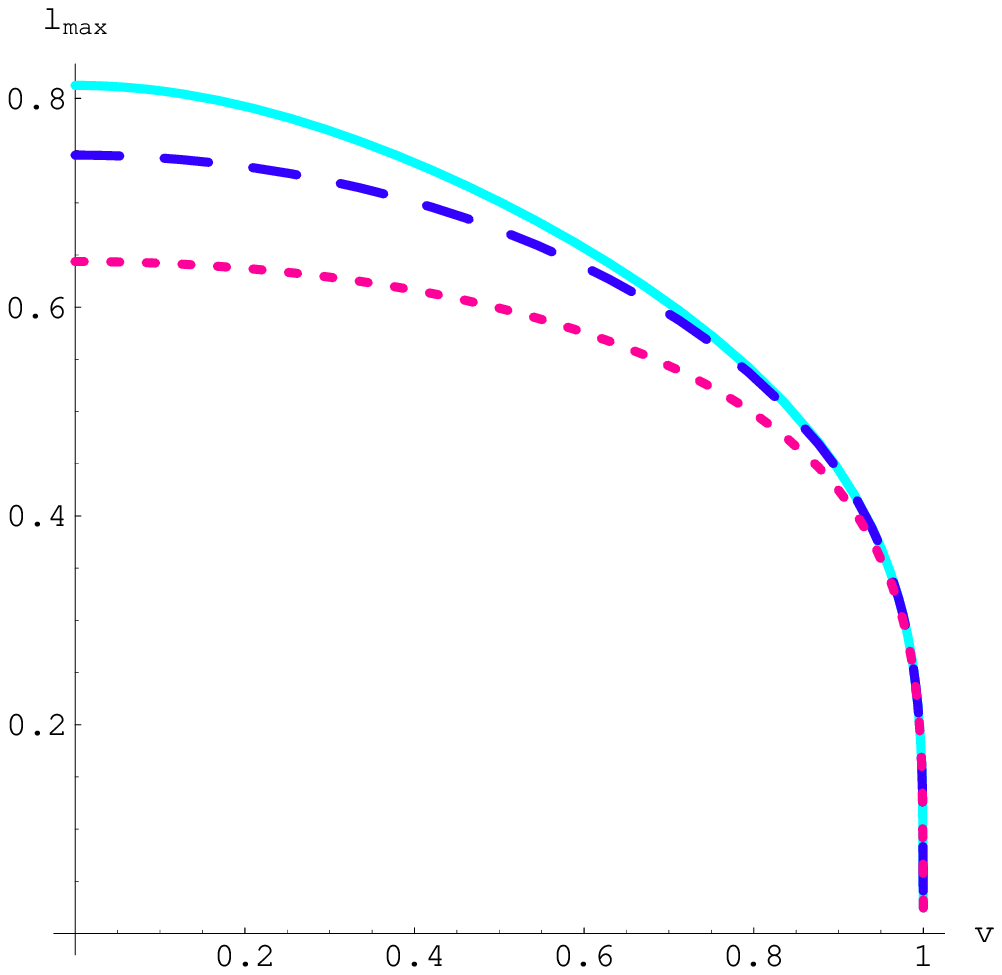} & \includegraphics[height=4.8cm]{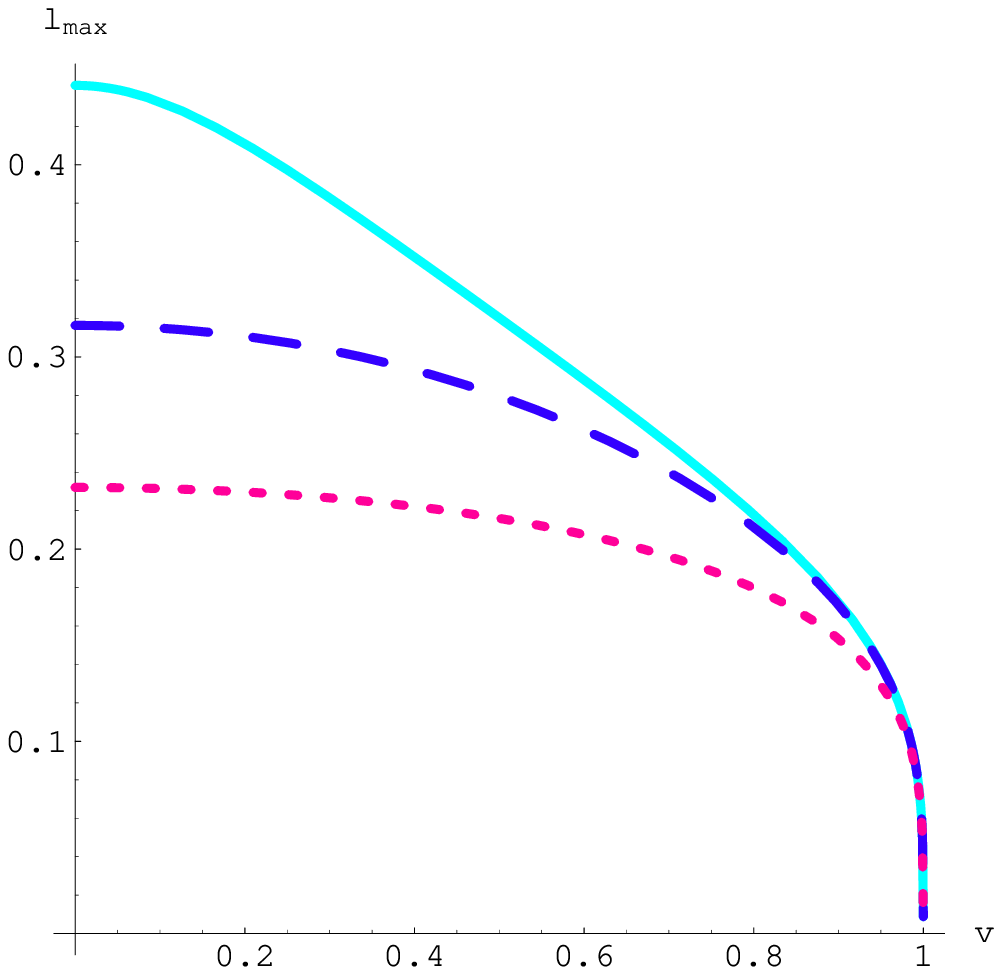}
& \includegraphics [height=4.8cm]{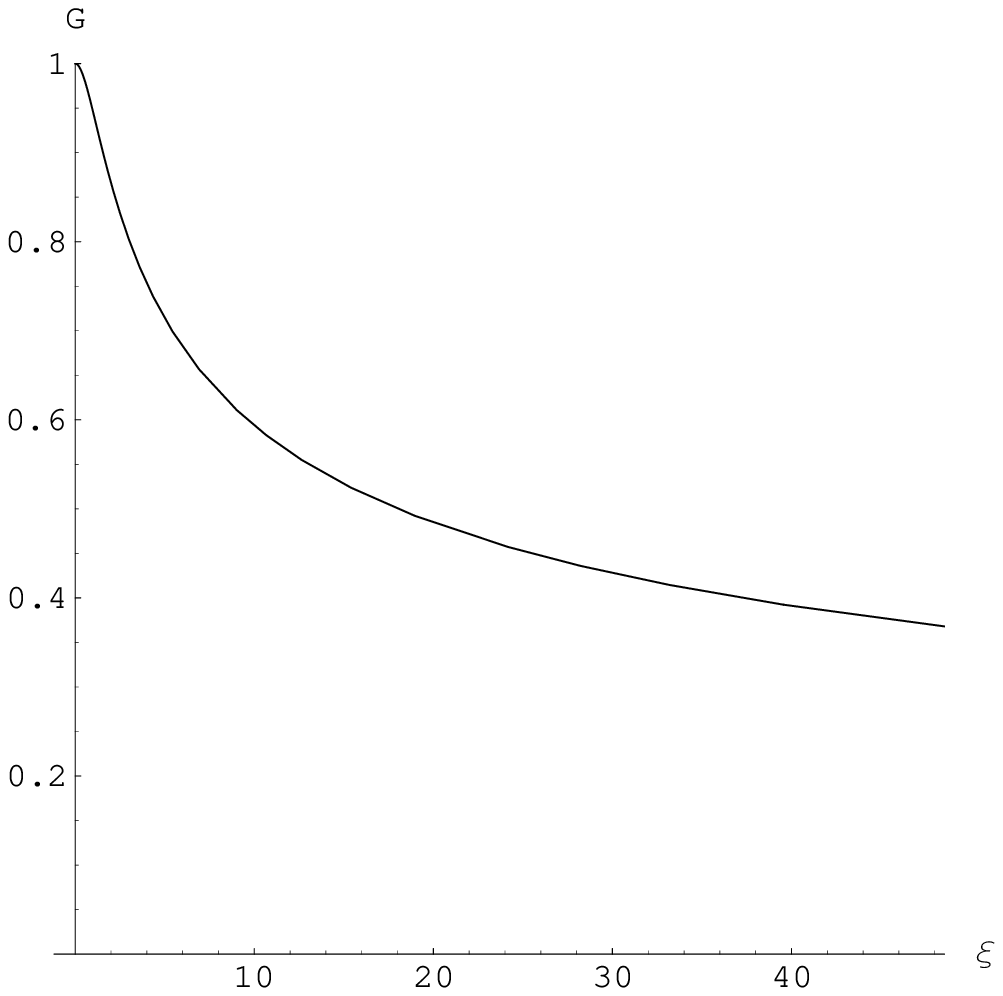}\\
(a) & (b) & (c)
\end{tabular}
\end{center}
\vskip -.5 cm \caption{(a), (b) Plots of $\ell_{\rm max}(v)$ for the case
of two angular momenta at $\th={\pi \ov 2}$ (with similar plots for $\th=0$)
with $\l=0.5$ and $\l=0.95$ respectively; The solid, dashed
and dotted lines are as in Fig. 2(c). (c) Plot of the function $G(\xi)$ for the case
of two angular momenta.}
\end{figure}

\no $\bullet$ For the case with $\th=\pi/2$, we have the
expression
\ba
\ell(v,\xi,y_0)\ = \
G(\xi)\ {\sqrt{y_0^4 - \g^2}\ov y_0^3} \
\int_{1}^\infty {dz \ov \sqrt{ ( z^2 - 1/y_0^2) ( z - \l^2/y_0^2) (z^2-1)}}\ .
\label{ca4i}
\ea
Performing a similar analysis as before we find that the various
coefficients entering \eqn{jhf9} are given by
\be
c_1 = 0.2106\ ,\qq c_2 = 0.100\ ,\qq c_3= 0.0898 \ .
\ee
For $\cF(v,\xi)$ we give its numerical values in the four corners
in the $(v,\xi)$-plane
\be
\cF(0,0)= 1.06\ ,\qq \cF(0,\infty)= 2.11\ ,\qq \cF(1,0)= 1\ ,\qq
\cF(1,\infty)= 1\ ,
\ee
showing that it is a slowly-varying function except for the case of low velocity and extremely large R-charge. However, the region of the $(v,\xi)$ parameter space where these deviations from unity appear is very small, as can be easily verified by an explicit plot of $\cF(v,\xi)$.

\no In both cases the function $G(\xi)$ is given by
\be
G(\xi) = (1 + F^2)^{-1/2} = \sqrt{1-\l^2}\ .
\ee
It is plotted in Fig. 3(c) and it has a strong
dependence on the dimensionless R-charge density $\xi$.


\subsubsection{One nonzero angular momentum}

Now, the functions $f_y(u)$ and $g(u)$ are given by
\be
f_y(u) = {1\ov R^4} \left[ u^2(u^2 + r_0^2 \cos^2 \theta) - \g^2
\m^4 \right]\ , \quad g(u) = {u^2(u^2 + r_0^2 \cos^2 \theta) -
\g^2 \m^4 \ov u^2 (u^2 + r_0^2) - \m^4}\ ,
\ee
and again the angular variable $\th=0$ or $\th={\pi\ov 2}$ for
consistency. The length integrals are given below.

\no $\bullet$ For the case with $\th=0$ after a change of
variables we find that
\ba
\ell(v,\xi,y_0) &\! = \! &
G(\xi) {\sqrt{y_0^4 + \l^2 y_0^2 - \g^2}\ov y_0^3}
\nonumber\\
&& \phantom{} \times
\int_{1}^\infty {d z \ov \sqrt{ z (z-1) ( z^2 + \l^2 z/y_0^2- 1/y_0^4) (z + 1 + \l^2/y_0^2)}}\ .
\label{cas2}
\ea
An analysis similar to before gives for the numerical factors in \eqn{jhf9} that
\be
c_1 = 0.10525\ ,\qq c_2 = 0.100\ ,\qq c_3= -0.01507 \ .
\ee
The numerical values of $\cF(v,\xi)$ in the four corners
of the $(v,\xi)$-plane are
\be
\cF(0,0)= 1.06\ ,\qq \cF(0,1.33)= 1.05\ ,\qq \cF(1,0)= 1\ ,\qq \cF(1,1.33)=1\ ,
\ee
showing again that it is indeed a very slowly varying function.

\begin{figure}[t]
\label{fig4}
\begin{center}
\begin{tabular}{ccc}
\includegraphics [height=4.8cm]{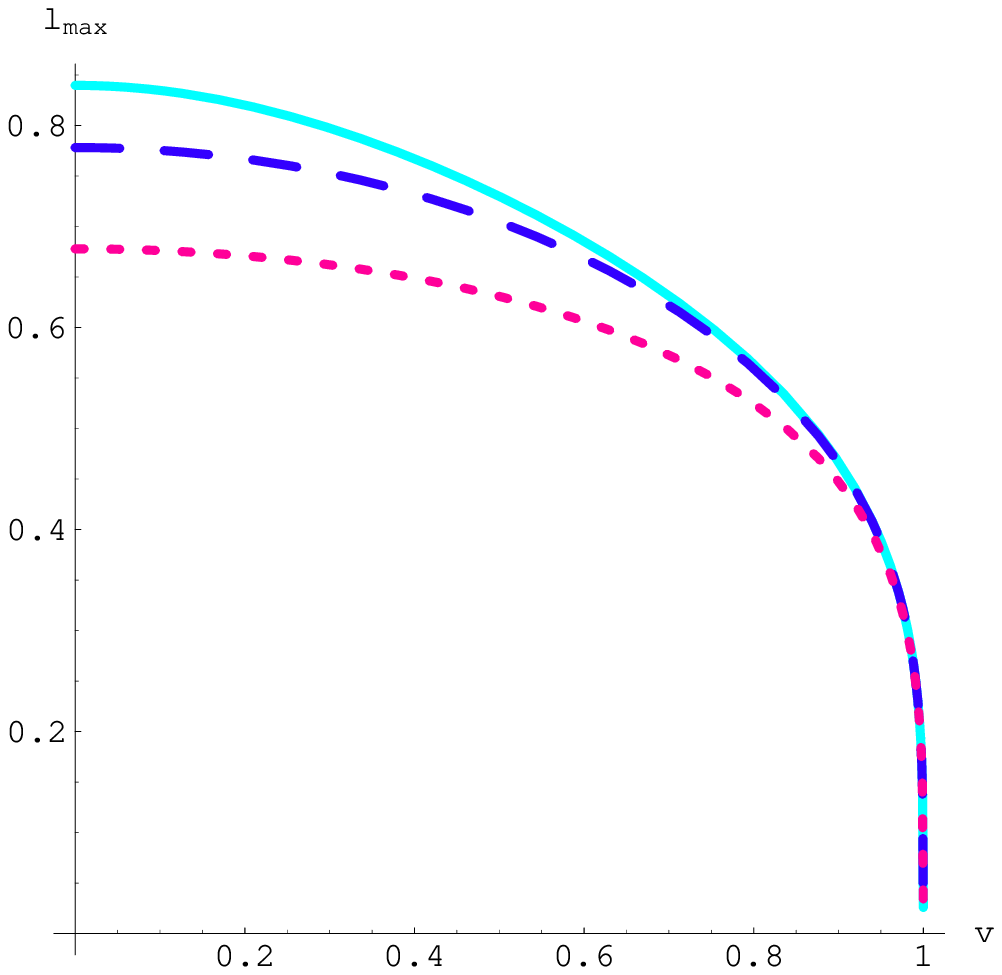} & \includegraphics[height=4.8cm]{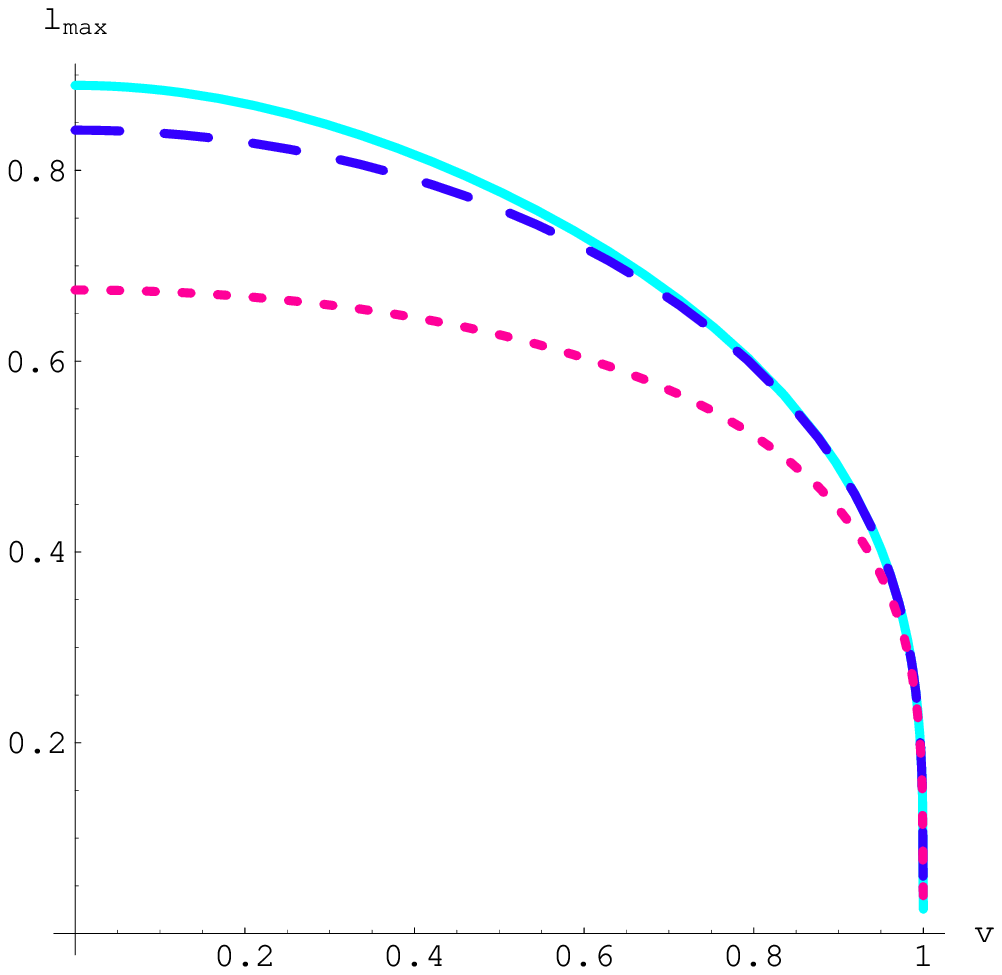}
& \includegraphics [height=4.8cm]{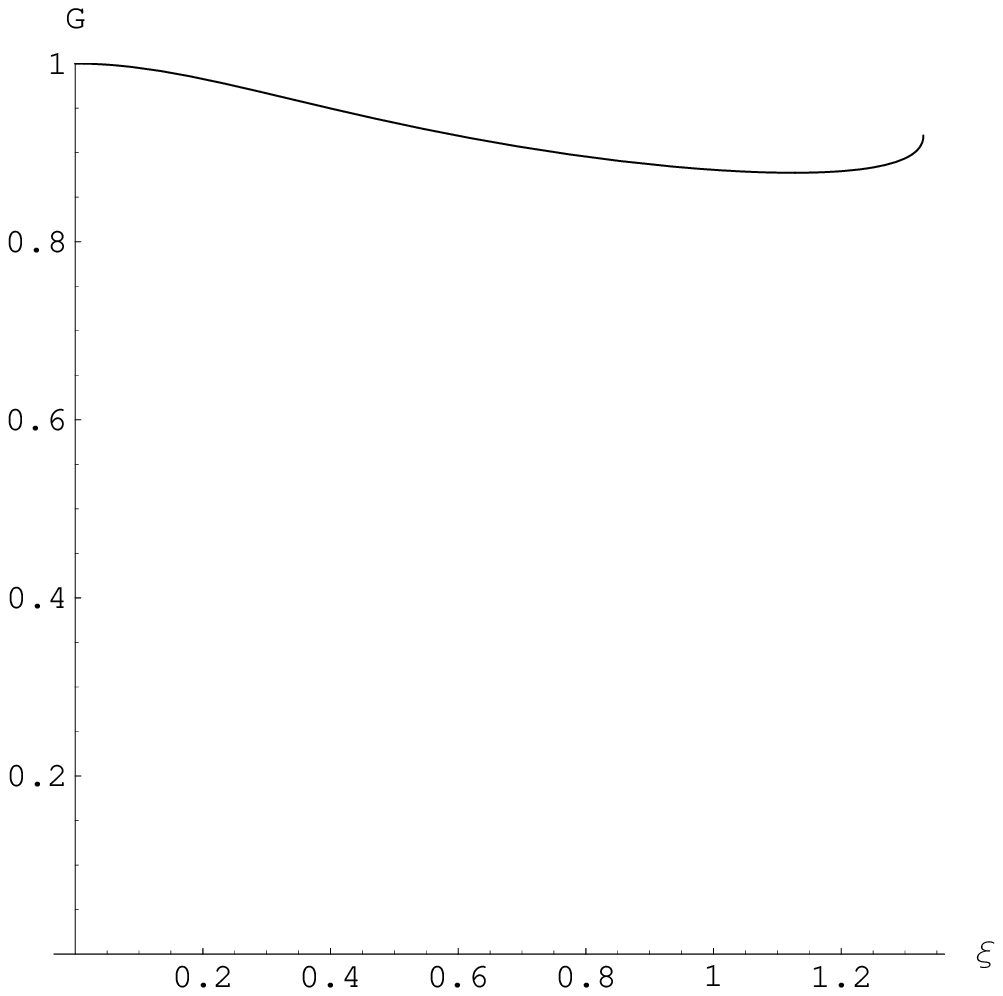}\\
(a) & (b) & (c)
\end{tabular}
\end{center}
\vskip -.5 cm \caption{(a), (b) Plots of $\ell_{\rm max}(v)$ for the case
of one angular momentum at $\th=0$ (with similar plots for $\th={\pi\ov 2}$)
with $\l=0.7$ and $\l=1.4$ respectively; the solid, dashed
and dotted lines are as in Fig. 2(c).
(c) Plot of the function $G(\xi)$ for the case
of one angular momentum.}
\end{figure}

\no $\bullet$ For the case with $\th=\pi/2$ we obtain after a
similar variable change that
\be
\ell(v,\xi,y_0)=
G(\xi) {\sqrt{y_0^4 - \g^2}\ov y_0^3}
\int_{1}^\infty {d z\ov \sqrt{z (z^2-1)( z^2 + \l^2 z/y_0^2- 1/y_0^4)}}\ .
\label{cas1}
\ee
Now the numerical factor in \eqn{jhf9} are
\be
c_1 = -0.2106\ ,\qq c_2 = 0.100\ ,\qq c_3= 0.0898 \ ,
\ee
while the numerical values of $\cF(v,\xi)$ in the four corners
of the $(v,\xi)$-plane are
\be
\cF(0,0)= 1.06\ ,\qq \cF(0,1.33)= 0.74\ ,\qq \cF(1,0)= 1\ ,\qq \cF(1,1.33)=1\ ,
\ee
showing that it is a slowly-varying function except for the case of low velocity
with large R-charge when further corrections become important.

\no
Now we have for the function $G(\xi)$ the expression
\be
G(\xi) = F^{-3/4} (2-F)^{-1/4} = \sqrt{1 + {\l^4 \ov 4}}
\left(\sqrt{1 + {\l^4 \ov 4}} - {\l^2 \ov 2} \right)^{1/2}\ .
\ee
This is depicted in Fig. 4(c) and shows a very weak dependence on the dimensionless
R-charge density $\xi$.

\no
One comment is in order here regarding the case of parallel motion having $\Th=0$. In that case the
form of the maximal screening length is the same with slightly different coefficients and
function $\cF$. The function $G(\xi)$ is the same and is independent of the angle $\Th$,
a fact that facilitates the computation of average values of the maximal screening lengths
in the next section.

\section{Plasma rest frame and averages}

So far we have not considered the most general situation where the $q\bar{q}$-pair axis
makes an arbitrary angle $\Th$ with the velocity vector.
A detailed investigation of this problem can be carried out either by numerical
analysis of the full equations \eqn{3-10} or
by an approximate analytic method that we will present here.
We have already noted that the maximum separation length is a function of $\Th$
through the combinations $\cos^2\Th$ and $\sin^2\Th$. Since its values for
the extreme case
$\Th=0$ and $\Th={\pi/2}$, do not differ very much, it is a good approximation
for the screening length in the rest frame of the pair to write
\be
L_{\rm max, q\bar q}(\Th) = L_{\perp} \sin^2\Th + L_{||} \cos^2\Th\ ,
\label{hfo}
\ee
as it reduces to the above mentioned extreme cases, with the maximum screening
lengths denoted by $L_{\perp}$ and $L_{||}$, in an obvious notation.
It is of interest to know the average value of this length over all angles.
To do that we have to know the probability for a $q\bar q$ pair to be moving
at an angle, formed by its axis and its velocity vector, between $\Th$ and $\Th+d\Th$.
The most reasonable assumption is that this probability is uniform, but in the plasma rest
frame. Denoting in the plasma rest frame by $\Th'$ the angle corresponding to the angle $\Th$ in
the $q\bar q$ rest frame, we have, due to length contraction in the direction $x$
parallel to the velocity vector, the relation
\be
\tan\Th' = \g\tan\Th \ ,
\ee
that is the length is not only contracted but also rotated.
From these the useful relations
\be
\cos^2\Th' = {\g^{-2} \cos^2\Th\ov \g^{-2} \cos^2\Th + \sin^2\Th}\ ,\qq
\sin^2\Th' = { \sin^2\Th\ov \g^{-2} \cos^2\Th + \sin^2\Th}
\ee
and their inverses
\be
\cos^2\Th = {\g^{2} \cos^2\Th'\ov \g^{2} \cos^2\Th' + \sin^2\Th'}\ ,\qq
\sin^2\Th = { \sin^2\Th'\ov \g^{2} \cos^2\Th' + \sin^2\Th'}\ ,
\ee
follow.
Then the probability distribution is
\be
P_{\rm plasma}(\Th')d\Th' = {2\ov \pi} d\Th' = {2\ov \pi \g}
{d\Th\ov \g^{-2} \cos^2\Th + \sin^2\Th} = P_{q\bar q}(\Th)d\Th\ .
\label{proobp}
\ee
Then the average value of the maximal screening length in the $q\bar q$-pair rest frame is
\ba
\bar L_{\rm max,q\bar q}  & = &   {2\ov \pi \g} \int_0^{\pi/2} d\Th\
{L_{\perp} \sin^2\Th + L_{||} \cos^2\Th\ov \g^{-2} \cos^2\Th + \sin^2\Th}
\nonumber\\
&  = &
{L_{\perp} + \g  L_{||}\ov 1+\g}\ =\ L_\parallel
+ (L_\perp -L_\parallel)\left ({1\ov \g} - {1\ov \g^2}\right) +
{\cal O}\left(1\ov \g^3\right) \ .
\label{dch1}
\ea
This is essentially of the form \eqn{jhf9}\footnote{There are two additional terms
appearing in the parenthesis of the form $1/\g$ and $\l^2/\g^2$.}
and for large enough $\g$ it is basically
controlled by the maximal screening length $L_\parallel$ along the pair's velocity vector.
The reason for this is the fact that the probability \eqn{proobp}, although uniform in the plasma
rest frame, it is peaked at $\Th=0$ with a width of order $1/\g$, in the $q \bar q$-pair rest frame.

\no
In the plasma rest frame the expression corresponding to \eqn{hfo}
differs from the above in that we have to take
into account the Lorentz contraction of the component of the separation length vector parallel
to the velocity. We find
\ba
L_{\rm max,plasma}(\Th') &  = &
  (L_{\perp} \sin^2\Th + L_{||} \cos^2\Th) \sqrt{\g^{-2} \cos^2\Th + \sin^2\Th}
\nonumber\\
& = &
{L_{\perp} \sin^2\Th' + L_{||} \g^2 \cos^2\Th'\ov  (\g^{2} \cos^2\Th' + \sin^2\Th')^{3/2}}\ ,
\ea
with average
\ba
\bar L_{\rm max,plasma}&  = & {2\ov \pi \g} \int_0^{\pi/2} d\Th\
{L_{\perp} \sin^2\Th + L_{||} \cos^2\Th\ov \sqrt{\g^{-2} \cos^2\Th + \sin^2\Th}}
\nonumber\\
& = & {2\ov \pi \g} \left[L_\perp + L_\parallel \ln (4e^{-1} \g)\right] + {\cal O}\left(1\ov \g^3\right)\ .
\ea
We see that there is a $1/\g$ suppression of the maximal screening length, the reason being,
as explained before,
that in the pair's rest frame the production is peaked near $\Th=0$. The presence of the
$\ln \g$ term in the overall coefficient is due to the contribution of the integral in
the region near $\Th=0$.
The last expression in the basis of our proposal for a maximum screening length in the
plasma rest frame in \eqn{jhf9abnn}.

\no
We note for completeness that, if we take the angle distribution to
be uniform in the $q\bar{q}$-pair rest frame, then the average value of
the maximum length in the rest frame of the pair
would be, instead of \eqn{dch1}
\be
\bar L_{\rm max,q\bar q} = { L_{\perp} + L_{||} \ov 2}\ ,
\ee
again of the form \eqn{jhf9}. Also the corresponding average in the plasma rest frame would be
\ba
\bar L_{\rm max,plasma}&  = & {2\ov \pi} \int_0^{\pi/2} d\Th\
(L_{\perp} \sin^2\Th + L_{||} \cos^2\Th) \sqrt{\g^{-2} \cos^2\Th + \sin^2\Th}
\nonumber\\
& = & \ha \left[L_\perp\ {}_2F\!_1(-\ha, {1\ov 2},2,v^2) +  L_\parallel\
{}_2F\!_1(-\ha, {3\ov 2},2,v^2)\right]
\\
& = & {2\ov 3\pi} (2 L_\perp + L_\parallel) + {1\ov 6\pi \g^2} \Big(2 L_\perp
- 5 L_\parallel + 6 L_\parallel \ln (4 \g)\Big) + {\cal O}\left(1\ov \g^4\right)\ .
\nonumber
\ea
The last expression is the basis of our second possibility, less likely in our opinion,
for a maximum screening length in the plasma rest
frame in \eqn{jhf9lab}.

\vskip .8 cm

\centerline{ \bf Acknowledgments}

\no
K.~S. would like to thank U. Wiedemann for a useful
discussion, as well as the Theory-Division at CERN and the
Institute of Physics of the University of Neuch\^atel for
hospitality and financial support during a considerable part of
this research.\hfill\break K.~S. also acknowledges the support
provided through the European Community's program ``Constituents,
Fundamental Forces and Symmetries of the Universe'' with contract
MRTN-CT-2004-005104, the INTAS contract 03-51-6346 ``Strings,
branes and higher-spin gauge fields'', the Greek Ministry of
Education programs $\rm \P Y\Th A\G OPA\S$ with contract 89194 and
the program $\rm E\Pi A N$ with code-number B.545.

\end{document}